\documentclass[12pt]{article}
\usepackage{amsfonts}
\usepackage{latexsym}
\usepackage{amsmath,amssymb}
\usepackage{verbatim}
\usepackage{setspace}
\usepackage{color}
\usepackage{physics}
\usepackage{tikz}
\usepackage{mathdots}
\usepackage{yhmath}
\usepackage{cancel}
\usepackage{color}
\usepackage{siunitx}
\usepackage{mathrsfs}  
\usepackage{array}
\usepackage{multirow}
\usepackage{amssymb}
\usepackage{gensymb}
\usepackage{tabularx}
\usepackage{hyperref}
\usepackage{booktabs}
\numberwithin{figure}{section}
\usetikzlibrary{fadings}
\usetikzlibrary{patterns}
\usetikzlibrary{shadows.blur}
\usetikzlibrary{shapes}

\usetikzlibrary{decorations.pathmorphing}

\tikzset{snake it/.style={decorate, decoration=snake}}

\usepackage[textheight=9in, textwidth=6.5in, letterpaper]{geometry}

\numberwithin{equation}{section}

\def\ip{${\mathcal I}^+$}

 \def\p{\partial}
 
 \def\bz{{\bar z}}
 
\def\0{{(0)}}
\def\1{{(1)}}
\def\2{{(2)}}

\def\n{\nabla}

\def\<{\langle }
\def\>{\rangle }
\def\[{\left[}
\def\]{\right]}
\def\bw{{\bar w}}

\newcommand{\bea}{\begin{eqnarray}}
\newcommand{\eea}{\end{eqnarray}}
\newcommand{\be}{\begin{equation}}
\newcommand{\ee}{\end{equation}}
\newcommand{\ba}{\begin{align}}
\newcommand{\ea}{\end{align}}

\renewcommand{\O}{\mathcal{O}}

\renewcommand{\epsilon}{\varepsilon}

   \makeatletter
  \let\over=\@@over \let\overwithdelims=\@@overwithdelims
  \let\atop=\@@atop \let\atopwithdelims=\@@atopwithdelims
  \let\above=\@@above \let\abovewithdelims=\@@abovewithdelims
\renewcommand\section{\@startsection {section}{1}{\z@}%
                                   {-3.5ex \@plus -1ex \@minus -.2ex}
                                   {2.3ex \@plus.2ex}%
                                   {\normalfont\large\bfseries}}

\renewcommand\subsection{\@startsection{subsection}{2}{\z@}%
                                     {-3.25ex\@plus -1ex \@minus -.2ex}%
                                     {1.5ex \@plus .2ex}%
                                     {\normalfont\bfseries}}

\begin{document}
\unitlength = 1mm
\ \\
\vskip1cm
\begin{center}

{ \LARGE {\textsc{Celestial Amplitudes as AdS-Witten Diagrams}}}

\vspace{0.8cm}
Eduardo Casali,
Walker Melton,
and Andrew Strominger

\vspace{1cm}

{\it  Center for the Fundamental Laws of Nature, Harvard University,\\
Cambridge, MA 02138, USA}

\begin{abstract}

Both celestial and momentum space amplitudes in four dimensions  are beset by divergences resulting from spacetime  translation and sometimes scale  invariance. In this paper we consider a (linearized) marginal deformation of the celestial CFT for 
Yang-Mills theory which preserves 2D conformal invariance but breaks both spacetime translation and scale invariance and involves a chirally coupled massive scalar. The resulting MHV celestial amplitudes are completely finite (apart from the usual soft and collinear divergences and isolated poles in the sum of the weights) and  take the canonical CFT form. Moreover, we show they can be  simply rewritten in terms of AdS$_3$-Witten contact diagrams which evaluate to the well-known $D$-functions, thereby forging a direct connection between flat and AdS holography.  

 \end{abstract}
\vspace{0.5cm}

\vspace{1.0cm}

\end{center}

\pagestyle{empty}
\pagestyle{plain}
\newpage
\tableofcontents
\def\gzz{\gamma_{z\bz}}
\def\vx{{\vec x}}
\def\p{\partial}
\def\po{$\cal P_O$}
\def\cN{{\cal N}_\rho^2 }
\def\N{${\cal N}_\rho^2 ~~$}
\def\G{\Gamma}
\def\a{{\alpha}}
\def\b{{\beta}}
\def\g{\gamma}
\def\ch{{\cal H}^+}
\def\hf{{\cal H}}
\def\Im{\mathrm{Im\ }}
\def\bpd{\bar{\partial}}
\def\hbh{{\cal H}_{\rm BH}}
\def\hout{{\cal H}_{\rm OUT}}
\def\ss{\Sigma_S}
\def\D{{\rm \Delta}}
\def \bw {{\bar w}}
\def \bz {{\bar z}}
\def\cS{{\cal S}}
\def\l{\lambda }
\def\d{{\delta}}
\def\n{{\rm SC}}
\def\ip{{\rm cft}}
\def\RR{\mathbb{K}}
\def\i{i^\prime}
\def\A{\mathcal{A}}
\def\zb{\bar{z}}
\def\adz{AdS$_3/\mathbb{Z}$}
\def\sll{$SL(2, \mathbb{R})_L$}
\def\slr{$SL(2, \mathbb{R})_R$}
\pagenumbering{arabic}

\tableofcontents
\section{Introduction}

Celestial scattering amplitudes \cite{Pasterski:2016qvg} are ordinary scattering amplitudes written in a conformal primary basis of asymptotic particle states, often defined as Mellin transforms of momentum space amplitudes. 
In any asymptotically flat 4D quantum theory of gravity, celestial amplitudes transform covariantly under the full local conformal group acting on the celestial sphere \cite{Cachazo:2014fwa,Kapec:2014opa}. This raises the hope  that decades of powerful results from the study of CFT$_2$ can be applied to constrain and compute amplitudes  in real-world 4D quantum gravity.  

Although much progress has been made in applying CFT$_2$ techniques to celestial amplitudes (see the recent reviews 
\cite{Raclariu:2021zjz,Pasterski:2021rjz,Pasterski:2021raf}),  this hope has been in part impeded by the fact that the 3- and 4- point celestial amplitudes have kinematic singularities which are direct consequences of translation invariance and, in the Yang-Mills (YM) case considered here, spacetime scale invariance. While fullly consistent with 2D conformal invariance, and obvious from the bulk perspective, these singularities are unfamiliar from the CFT$_2$ perspective. 

In this paper we eliminate all of these singularities by the chiral coupling\footnote{Of the form $\phi \tr (F^-)^2$ where $F^-$ is the anti-self-dual part of the gauge field.} of a massive scalar. We  expand the amplitudes, to linear order, around a classical solution in which only the scalar is nonzero. This solution explicitly breaks translation and scale invariance but is judicously chosen so as not to break the 
2D conformal invariance. From the point of view of the boundary celestial CFT (CCFT) it is a marginal dimension $(1,1)$ operator. The result is that all three and higher point amplitudes are generically smooth and finite and take the standard CFT$_2$ form.\footnote{Since the smooth linear deformation of the amplitudes depend on the same number of variables as the singular undeformed ones, it is possible and interesting to ask if the latter can be reconstructed from the former.}

Our results were informed by the inspiring recent papers of Costello  and Paquette \cite{Costello:2022wso} and of Fan, Fotopolous, Stieberger, Taylor and Zhu \cite{Fan:2022vbz}. We follow these authors in  using scalars to break translation invariance, but the detailed constructions differ.

A significant byproduct of our investigations is the discovery of a simple method to express 
the 4D scattering amplitudes in terms of Witten diagrams in AdS$_3$. The basic idea is to write space as a foliation of hyperbolae labelled by a parameter $\tau$ and then do the $\tau $ integral first.  For celestial MHV diagrams the integral can be deformed to a closed contour in the complex $\tau$ plane whose residues are  proportional to scalar 
contact Witten diagrams. These have been the subject of much study $e.g.$~\cite{Chen:2017gwd, Hijano:2016DFunction} and are known as $D$-functions.  More generally beyond MHV  loop and exchange diagrams will appear. 
This forges a direct connection between 4D celestial holography and AdS$_3$/CFT$_2$.

In this paper we consider only a limited class of examples but, from the form of our construction,  it appears plausible that the connection between celestial amplitudes and Witten diagrams is a quite general and potentially fruitful one.  A universal construction would be of great interest.

This paper is organized as follows. Section 2 contains a basic review of Klein space and celestial amplitudes and establishes conventions.  In section 3 we introduce the theory of interest:
Yang-Mills in Klein space with a complex massive scalar coupled to the square of the anti-self-dual curvature $F^-$. In section 4 we compute  the 3 and 4 point amplitudes to linear order in the perturbation. Since the leading, scalar-independent term vanishes everywhere off of a singular locus, at generic points the first order correction is the leading term in the amplitude. 
In section 5, we examine soft and collinear limits of the 4 point function, and find that they paint a coherent picture agreeing  with expectation.

\section{Preliminaries}

In this section we review conventions and equations for celestial amplitudes in Klein space. We also collect some results on OPEs and soft theorems in order to make the paper more self-contained. The reader familiar with these topics can safely skip to the next section.

\subsection{The Structure of Klein Space}
\begin{figure}[h!]

	\begin{center}
\tikzset{every picture/.style={line width=0.75pt}} 

\begin{tikzpicture}[x=0.75pt,y=0.75pt,yscale=-1,xscale=1]

\draw  [fill={rgb, 255:red, 221; green, 221; blue, 221 }  ,fill opacity=1 ] (30,30) -- (210,210) -- (30,210) -- cycle ;
\draw    (30,210) -- (120,120) ;

\draw (25,9) node [anchor=north west][inner sep=0.75pt]    {$i'$};
\draw (213,198) node [anchor=north west][inner sep=0.75pt]    {$i^{0}$};
\draw (120,124) node [anchor=north west][inner sep=0.75pt]   [align=left] {$ $};
\draw (119,99) node [anchor=north west][inner sep=0.75pt]    {$\mathcal{I}$};
\draw (44,109) node [anchor=north west][inner sep=0.75pt]    {$\tau \in \mathbb{R}$};
\draw (94,169) node [anchor=north west][inner sep=0.75pt]    {$\tau \in i\mathbb{R}$};

\end{tikzpicture}
	\end{center}
	\caption{The toric Penrose diagram for signature $(2,2)$ Klein space. A Lorentzian torus is fibered over every point on the diagram, except the left vertical edge where one cycle degenerates to zero length and the bottom edge where the other cycle degenerates.  Null geodesics at a $i^0$ labels spatial infinity and $i'$ timelike infinity. $\mathcal{I}$ labels null infinity, which is a Lorentzian torus fibered over a null interval. The spacetime divides into an upper and lower wedge characterized by the sign of $\tau^2=-x^\mu x_\mu$. \label{fig:1}}
	\end{figure}
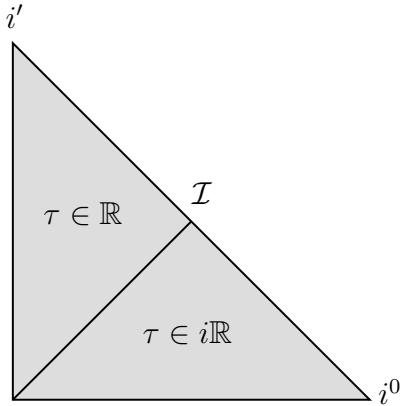

Modern treatments of scattering amplitudes are implicitly or explicitly performed in (2,2) signature Klein space, in that Lorentz spinors and their conjugates are varied independently. Here we shall explicitly work in  Klein space, where celestial amplitudes take the form of a correlations in a Lorentzian CFT on the  celestial torus,  where certain calculations are simplified \cite{Atanasov:2021,Atanasov:2021CPW}. The metric of Klein space is 
\begin{equation}
    ds^2 = -dx_0^2 + dx_1^2 - dx_2^2 + dx_3^2.
\end{equation}
Unlike (3,1)-signature Minkowski space, Klein space has a single null infinity and has the toric Penrose structure shown in Figure \ref{fig:1} \cite{Atanasov:2021}.
In Klein space, null infinity takes the form of a Lorentzian torus fibered over a null interval, while timelike infinity $i'$ and spacelike infinity $i^0$ are geometrically AdS$_3/\mathbb{Z}$ \cite{Atanasov:2021}.

A massless momentum in (2,2)-signature Klein space $p^\mu$ can be parametrized by a positive frequency $\omega$, an in/out label $\epsilon=\pm1$ and a coordinate on (a diamond patch of) the celestial torus $z,\zb \in \mathbb{R}$
\begin{equation}
\begin{split}
    p^\mu &= \epsilon\omega q^\mu(z,\zb),\ \omega > 0 \\
   q^\mu(z,\zb) &=  (1+z\zb,z+\zb,z-\zb,1-z\zb)
  \end{split}
\end{equation}
Under $\mathrm{SO}(2,2) \simeq \frac{\mathrm{SL}_2(\mathbb{R})\times \mathrm{SL}_2(\mathbb{R})}{\mathbb{Z}_2}$ transformations, $z, \zb$ transform by Moebius transformations. In this paper they are independent real variables; under some continuations to Minkowski space they become complex conjugates (hence the notation). While there is no sharp distinction between ingoing and outgoing particles in Klein space, the null momenta $\omega(1+z\zb,z+\zb,z-\zb,1-z\zb)$ for $\omega > 0$ cover only half of the space of null momenta; on the celestial torus, the coordinates $z, \zb$ cover a single diamond  patch that composes half of the full celestial torus, and the ingoing/outgoing label $\epsilon = \pm 1$ tells us towards which patch of the celestial torus the null momenta $q^\mu$ points. The $\mathrm{SO}(2,2)$ symmetry guarantees that there is ultimatley no preferred diamond patch, and the decomposition into in and out modes is a coordinate choice, albeit a useful one in comparing to Minkowskian formulae.  
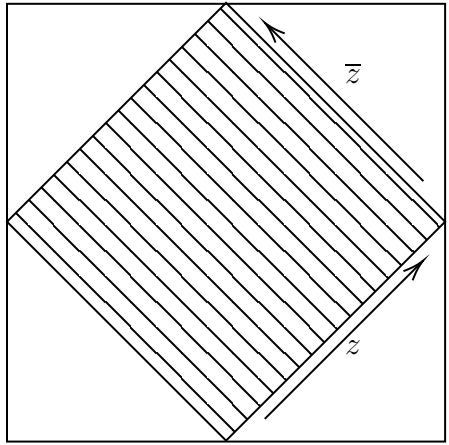
\begin{figure}[h!]\label{fig:2}
\begin{center}

 
\tikzset{
pattern size/.store in=\mcSize, 
pattern size = 5pt,
pattern thickness/.store in=\mcThickness, 
pattern thickness = 0.3pt,
pattern radius/.store in=\mcRadius, 
pattern radius = 1pt}
\makeatletter
\pgfutil@ifundefined{pgf@pattern@name@_bempymz16}{
\pgfdeclarepatternformonly[\mcThickness,\mcSize]{_bempymz16}
{\pgfqpoint{0pt}{-\mcThickness}}
{\pgfpoint{\mcSize}{\mcSize}}
{\pgfpoint{\mcSize}{\mcSize}}
{
\pgfsetcolor{\tikz@pattern@color}
\pgfsetlinewidth{\mcThickness}
\pgfpathmoveto{\pgfqpoint{0pt}{\mcSize}}
\pgfpathlineto{\pgfpoint{\mcSize+\mcThickness}{-\mcThickness}}
\pgfusepath{stroke}
}}
\makeatother
\tikzset{every picture/.style={line width=0.75pt}} 

\begin{tikzpicture}[x=0.75pt,y=0.75pt,yscale=-1,xscale=1]

\draw  [pattern=_bempymz16,pattern size=9.675pt,pattern thickness=0.75pt,pattern radius=0pt, pattern color={rgb, 255:red, 0; green, 0; blue, 0}] (120.48,10) -- (230.96,120.48) -- (120.48,230.96) -- (10,120.48) -- cycle ;
\draw   (10,10.48) -- (230.96,10.48) -- (230.96,231.43) -- (10,231.43) -- cycle ;
\draw    (140,220) -- (218.59,141.41) ;
\draw [shift={(220,140)}, rotate = 135] [color={rgb, 255:red, 0; green, 0; blue, 0 }  ][line width=0.75]    (10.93,-3.29) .. controls (6.95,-1.4) and (3.31,-0.3) .. (0,0) .. controls (3.31,0.3) and (6.95,1.4) .. (10.93,3.29)   ;
\draw    (220,100) -- (141.41,21.41) ;
\draw [shift={(140,20)}, rotate = 45] [color={rgb, 255:red, 0; green, 0; blue, 0 }  ][line width=0.75]    (10.93,-3.29) .. controls (6.95,-1.4) and (3.31,-0.3) .. (0,0) .. controls (3.31,0.3) and (6.95,1.4) .. (10.93,3.29)   ;

\draw (179,179) node [anchor=north west][inner sep=0.75pt]    {$z$};
\draw (179,39) node [anchor=north west][inner sep=0.75pt]    {$\overline{z}$};

\end{tikzpicture}

\end{center}
\caption{The celestial torus. The top/botttom and left/right edges of the square are identified. Null momenta with $z,\bar{z}, \omega > 0, \epsilon = 1$ cover a single Poincaré patch (shaded). Momenta with $\epsilon = -1$ cover the unshaded Poincaré patch}
\end{figure}

Using the basis for the Kleinian Pauli matrices 
\begin{equation}
    \sigma_\mu = \left(\begin{bmatrix} 1 & 0 \\ 0 & 1 \end{bmatrix}, \begin{bmatrix} 0  & 1 \\ 1 & 0 \end{bmatrix}, \begin{bmatrix}0 & 1 \\ -1 & 0 \end{bmatrix}, \begin{bmatrix} -1 & 0 \\ 0 & 1 \end{bmatrix}\right)
\end{equation}
we can identify $p_{\alpha\dot{\alpha}} = (p \cdot \sigma)_{\alpha\dot{\alpha}} = \lambda_\alpha\tilde{\lambda}_{\dot{\alpha}}$ where 
\begin{equation}
    \lambda_\alpha = \epsilon\sqrt{2\omega}\begin{bmatrix} z \\ 1 \end{bmatrix},\ \tilde{\lambda}_{\dot{\alpha}} = \sqrt{2\omega}\begin{bmatrix} \zb \\ 1 \end{bmatrix}.
\end{equation}
Angle and square brackets then take the form $\langle ij\rangle = 2\epsilon_i\epsilon_j\sqrt{\omega_i\omega_j}z_{ij},\ [ij] = 2\sqrt{\omega_i\omega_j}\zb_{ij}$. 

Working explicitly in Klein space can simplify integrals over the bulk because the timelike and spacelike constant-$\tau$ slices are both geometrically AdS$_3/\mathbb{Z}$. We can cover the entirety of Klein space with coordinates $(\tau,\hat{x}_\pm)$ where $\hat{x}_\pm^2 = \mp 1$, such that  in the timelike region 
\begin{equation}
ds^2 = -d\tau^2 + \tau^2d^2H_3 
\end{equation}
and in the spacelike region 
\begin{equation}
ds^2 = d\tau^2-\tau^2d^2H_3
\end{equation}
where $d^2H_3$ is a metric for (2,1)-signature AdS$_3/\mathbb{Z}$ and $\tau$ runs from $0$ to $\infty$ along the positive real axes in both the spacelike and timelike regions. 

For sufficiently smooth functions $f(x)$, we can rotate the $\hat{x}_-$ contour and evaluate the integral over Klein space as an integral over $\tau$ running along the positive imaginary and then positive real axis, and an integral over a timelike unit vector $\hat{x}$. 
\begin{equation}
    \int d^4xf(x) \to \int_{i\infty}^\infty d\tau\tau^3\int \widetilde{d^3\hat{x}}f(\tau\hat{x}).
\end{equation}
where $\widetilde{d^3\hat{x}}$ is the volume element and $\hat x$ the coordinate on AdS$_3/\mathbb{Z}$. This contour rotation is valid provided that $f$ falls off sufficiently quickly on the AdS$_3/\mathbb{Z}$ slices and does not have poles in a particular region of the $\hat{x}$ integral. A suitable choice of $i\epsilon$ procedure guarantees that this contour rotation is valid for our integrand. 

\subsection{Celestial Amplitudes for Klein-Space Amplitudes}
  
Given a momentum-space amplitude $\mathcal{A}(q_1,\ldots,q_n)$ (including the momentum-conserving $\delta$ function), we can transform the amplitude to the celestial sphere by Mellin transforming with respect to the frequency $\omega$:
\begin{equation}
    \mathcal{A}(1^{\epsilon_1}_{\Delta_1,a_1}\cdots n^{\epsilon_n}_{\Delta_n,a_n}) = \prod_{j=1}^n\int_0^\infty d\omega_j\omega_j^{\Delta_j-1} \mathcal{A}_{a_1\ldots a_n}(\epsilon_1\omega_1q(z_1,\zb_1),\ldots, \epsilon_n\omega_nq(z_n,\zb_n)).
\end{equation}
where $a_i$ label properties of the particle such as spin or color. Under $\mathrm{SO}(2,2)$ transformations, the resulting celestial amplitudes transform as correlation functions of conformal primary operators of weight $\Delta_j$ and spin $J = s_j$, where $s_j$ is the spin of the $j$-th particle. 

    \subsection{Celestial OPEs and Conformally Soft Theorems}
In a CFT, higher-point amplitudes can be built up from lower point amplitudes using the operator product expansion. 
Given a set of correlation functions, we can investigate the OPE structure by looking at the limit $z_{12} \to 0$. For celestial amplitudes, this describes the scattering of collinear gluons, and transforming known collinear expansions of color-ordered gluon amplitudes gives the leading order  celestial OPEs \cite{Pate:2019ope, Fan:2019ope}
\begin{equation}
\begin{split}
    \A(1^+_{\Delta_1,+}2^+_{\Delta_2,+}\cdots n^{\epsilon_n}_{\Delta_n,s_n}) &\sim \frac{1}{2z_{12}}\sum_{m=0}^\infty\frac{B(\Delta_1+m-1,\Delta_2-1)}{m!}\zb_{12}^m\bpd^m\A(2^+_{\Delta_1+\Delta_2-1,+}\cdots n^{\epsilon_n}_{\Delta_n,s_n}) \\
    \A(1^+_{\Delta_1,-}2^+_{\Delta_2,+}\cdots n^{\epsilon_n}_{\Delta_n,s_n}) &\sim \frac{1}{2z_{12}}B(\Delta_1+1,\Delta_2-1)\A(2^+_{\Delta_2+\Delta_3-1,-}\cdots n^{\epsilon_n}_{\Delta_n,s_n}) 
\end{split}
\end{equation}
For full amplitudes, these OPEs pick up structure-constant prefactors.

Even though celestial amplitudes describe scattering of particles whose wavefunction contains modes of arbitrarily high energy, momentum-space soft theorems have an analog in the case of celestial amplitudes. These are dubbed conformal soft theorems \cite{Pate:2019soft} and are obtained in the limit when one of the conformal dimensions approaches $1$. For Yang-Mills, the leading conformally soft theorem tells us that, in color-ordered amplitudes,
\begin{equation}
\begin{split}
    \lim_{\Delta_1\to 1}(\Delta_1-1)\A(1^{\epsilon_1}_{\Delta_1,+}2^{\epsilon_2}_{\Delta_2,s_2} \cdots n^{\epsilon_n}_{\Delta_n,s_n}) &= \frac{1}{2}\frac{z_{n2}}{z_{n1}z_{12}}\A(2^{\epsilon_2}_{\Delta_2,s_2}\cdots n^{\epsilon_n}_{\Delta_n,s_n}).
\end{split}
\end{equation}
There is in fact a tower of conformally soft theorems generated by soft modes obtained from the residues of celestial amplitudes when one of the weights becomes an integer less than or equal to one. These soft modes were shown to generate interesting algebras in tree-level gravity and gauge theory, as well as in self-dual gravity \cite{Guevara:2021salg, Strominger:2021mtt, Ball:2021tmb,Adamo:2021lrv}.

\section{The Yang-Mills and Massive Scalar Theory}

We motivate our celestial amplitude from a particular scalar-gluon amplitude in a Yang-Mills theory coupled to a massive complex scalar. The Lagrangian of this theory is
\begin{equation}\label{YM_dil_L}
\mathcal{L}= \frac{1}{2}\partial_\mu\phi\partial^\mu\phi^* -\frac{m^2}{2}|\phi|^2 -\frac{1}{4}\tr F_{\mu\nu}F^{\mu\nu} -\frac{1}{4}\phi\tr F^-_{\mu\nu}F^{-\mu\nu}-\frac{1}{4}\phi^* F^+_{\mu\nu}F^{+\mu\nu}
\end{equation}
where 
$F^+$ is the self-dual curvature, $F^{-}$ the anti-self-dual curvature, and we set the gauge coupling to unity. In the case of a massless scalar, this Lagrangian was studied in \cite{Dixon:2004za} in relation to Higgs amplitudes, and more recently in \cite{Fan:2022vbz} where it was  applied to celestial amplitudes.\footnote{These authors actually study something slightly different with a complex $\phi$ with a source, but the computations are closely related.} The set up for our computation is similar in spirit to \cite{Fan:2022vbz} but differs in the way we couple the background to the amplitude and on the choice of scalar profile. It is also similar to the twisted holographic models of \cite{Costello:2022wso}, with a modified kinetic term which does not affect single-trace amplitudes. 

In this work we are interested in the single trace, tree-level, one-scalar and n-gluons amplitudes. In the case of a massless scalar, the 3- and 4-point color-ordered amplitudes were computed in \cite{Dixon:2004za} to be 
\begin{equation}\label{dil_4gluon}
\begin{split}
    \mathcal{A}_4(\phi,1^-,2^-,3^+) &=\frac{\langle 12\rangle^4}{\langle 12\rangle\langle 23\rangle\langle 34\rangle\langle 41\rangle}\delta^{(4)}\left(\sum_{i=1}^3p_i + Q\right) \\
    \mathcal{A}_5(\phi,1^-,2^-,3^+,4^+)&=\frac{\langle 12\rangle^4}{\langle 12\rangle\langle 23\rangle\langle 34\rangle\langle 41\rangle}\delta^4\left(\sum_{i=1}^4p_i+Q\right).
\end{split}
\end{equation}
where it was also conjectured that the pattern extends to higher multiplicity MHV amplitudes
\begin{equation}\label{dil_ngluon}
    \mathcal{A}_{n+1}(\phi,1^-,2^-,3^+,\cdots,n^+)=\frac{\langle 12\rangle^4 }{\langle 12\rangle \langle23\rangle\cdots\langle n1\rangle}\delta\left(\sum_{i=1}^n p_i+Q\right).
\end{equation}
These are the same as the usual n-point MHV gluon amplitudes, except that the momenta of the gluons sum to to the momentum of the scalar field $\sum_{i=1}^np_i=-Q$. Note that the simple form of this amplitude is due to the chiral way the scalar couples to the Yang-Mills theory. Since these amplitudes are rational functions of the external momenta their analytic continuation from Minkowski space to Klein space is trivial. At leading trace, the scalar field only participates as an external particle, i.e. it is not exchanged in any diagram contributing to this amplitude. Its propagator, along with its mass, never appears explicitly so the form of the massless and massive amplitudes are the same.

Given an on-shell background profile for the scalar $\phi(X)$ it can be expanded in a plane wave basis $\phi(X)=\int \widetilde{d^3Q}\,  e^{iQ\cdot X}\, g(Q)$ with Fourier coefficients $g(Q)$ with $\widetilde{d^3Q}$ is the invariant measure. The amplitudes \eqref{dil_ngluon} can then be coupled to a generic on-shell background by integrating over the scalar phase space
\begin{equation}\label{coupled_amp}
    \mathcal{A}^{\phi}_n(1^-,2^-,3^+,\cdots,n^+)=\int \widetilde{d^3Q} \,g(Q)\,\mathcal{A}_{n+1}(\phi,1^-,2^-,3^+,\cdots,n^+).
\end{equation}
These gluon amplitudes, coupled to a particular choice of background scalar, are what we'll Mellin transform to obtain the celestial amplitudes. 

\begin{figure}    \label{fig:dil_4pt}
     \centering
\tikzset{every picture/.style={line width=0.75pt}} 

\begin{tikzpicture}[x=0.75pt,y=0.75pt,yscale=-1,xscale=1]

\draw[snake it]    (220,40) -- (260,80) ;
\draw[snake it]    (260,80) -- (220,120) ;
\draw[snake it]    (310,80) -- (260,80) ;
\draw[snake it]   (350,40) -- (310,80) ;
\draw[snake it]    (350,120) -- (310,80) ;
\draw  [fill={rgb, 255:red, 246; green, 0; blue, 0 }  ,fill opacity=1 ] (262.06,85.14) .. controls (259.22,86.28) and (256,84.9) .. (254.86,82.06) .. controls (253.72,79.22) and (255.1,76) .. (257.94,74.86) .. controls (260.78,73.72) and (264,75.1) .. (265.14,77.94) .. controls (266.28,80.78) and (264.9,84) .. (262.06,85.14) -- cycle ;
\end{tikzpicture}
\caption{Contribution to 4-point gluon scattering around scalar background.}
\end{figure}
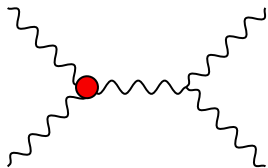

We can motivate our choice of coupling to the scalar background from the usual background field expansion. Treating the scalar as a fixed, classical, on-shell profile $\phi(X)$ consider the 4-point MHV amplitude in this background. It is easy to see from Feynman diagrams that contributions linear\footnote{In a background scalar the gauge field propagator is modified in a non-trivial way, but since we are only interested in the contributions linear in the background (these would be equivalent to computing the effect of a marginal deformation on the CCFT to first order) we treat the terms $\phi(F^-)^2$ as a coupling, leaving the gauge field with its canonical propagator.} in the background scalar always lead to one leftover integral over the insertion of the background. For example, the contribution from the diagram in figure \ref{fig:dil_4pt} is schematically given by
\begin{equation}
    \int d^4 X \int d^4 Y \,\phi(x)\,e^{i(p_1+p_2)\cdot X}G(X,Y)e^{i(p_3+p_4)\cdot Y}(\cdots)
\end{equation}
with $G(X,Y)$ the scalar propagator between the two interaction vertices and $(\cdots)$ the denominator corresponding to the particular graph.
Writing the propagator and scalar profile in momentum space and performing the integrals one obtains
\begin{equation}
   \int \widetilde{d^3Q}\, G(Q)\, \delta^4(\sum_{i=1}^4 p_i+Q)\frac{(\cdots)}{(p_1+p_2)^2}
\end{equation}
where the details of the particular diagram are contained in the terms to the right of the delta function. Since every diagram linear in the background is of this form, we reproduce the sum over Feynman diagrams resulting in the amplitudes \eqref{dil_ngluon} multiplied by an on-shell wavefunction $g(Q)$. Summing over all contributions under the same phase space integral recovers the definition of \eqref{coupled_amp} for the 4-point coupled amplitude. Note that in computing the full 4-point amplitude in this background there are other contributions to the amplitude coming solely from the $F^2$ term which are insensitive to the scalar background. Since we are only interested in the contributions that do couple to the background we don't discuss these terms further.

\section{Celestial MHV Amplitudes in a Scalar Background}

In this section we will compute the Mellin transform of \eqref{coupled_amp} in a particular background scalar profile which is dual to a marginal perturbation of the boundary CCFT. 
We will work to linear order in the perturbation. The perturbation  is a Bessel function\footnote{For the massless case  $\phi^\pm_{\Delta}(X|z,\zb) = \frac{(\mp i\epsilon)^\Delta\Gamma(\Delta)}{(-q(z,\zb) \cdot X \mp i\epsilon)^\Delta}$; the massive Bessel solutions can be found explicitly in \cite{Pasterski:2016qvg} and involves Bessel functions.}
\begin{equation}
    \phi_B(X) = \int d^2z\left[\phi_{\Delta=2,m}^+(X|z,\zb) + \phi_{\Delta=2,m}^-(X|z,\zb)\right] = \frac{8\pi^2}{im\tau}K_1(-im\tau).
\end{equation}
This is the unique solution to the massive wave equation (up to normalization) that depends only on $\tau = \sqrt{-X^2}$, and is exponentially damped for positive $\Im\tau$. Coupling the theory to a massive background field achieves two things: it breaks the tree-level scale invariance of Yang-Mills amplitudes, turning the $\delta(\beta)$ in front of conventional celestial Yang-Mills amplitudes into a smooth function of $\beta$; and it breaks translation invariance, removing the distributional support of low point amplitudes and repalcing it with more familiar CFT$_2$ expressions. Moreover, the dampening factor in $\Im\tau$ makes the celestial amplitudes well defined for generic $\mathrm{Re\ }\beta$.

Starting with the momentum space amplitude coupled to the background scalar field in equation \ref{coupled_amp}, its Mellin transform with respect to the external momenta gives
\begin{multline}
       \A(1^{\epsilon_1}_{\Delta_1,-}2^{\epsilon_2}_{\Delta_2,-}3^{\epsilon_3}_{\Delta_3,+}\cdots n^{\epsilon_n}_{\Delta_n,+}) \\=\frac{1}{2^{n-4}}\frac{z_{12}^3}{z_{23}\cdots z_{n1}} \int \widetilde{d^3Q}\,g(Q)d\omega_1\omega_1^{\Delta_1}d\omega_2\omega_2^{\Delta_2}\prod_{j=3}^nd\omega_j\omega_j^{\Delta_j-2} 
        \delta^{(4)}\left(Q + \sum_{j=1}^n \epsilon_j\omega_jq(z_j,\zb_j)\right) \\
    = \frac{1}{2^{n-4}}\frac{z_{12}^3}{z_{23}\cdots z_{n1}}\int \frac{d^4X}{(2\pi)^4}\int \widetilde{d^3Q}\,g(Q)  \int_0^\infty d\omega_1\omega_1^{\Delta_1}d\omega_2\omega_2^{\Delta_2}\prod_{j=3}^nd\omega_j\omega_j^{\Delta_j-2} e^{i\left(Q + \sum_j\epsilon_j\omega_jq(z_j,\zb_j)\right)\cdot X} \\
    = \frac{1}{2^{n-4}}\frac{z_{12}^3}{z_{23}\cdots z_{n1}}\int\frac{d^4X}{(2\pi)^4}\phi_B(X)\phi^{\epsilon_1}_{\Delta_1+1}(X|z_1,\zb_1)\phi^{\epsilon_2}_{\Delta_2+1}(X|z_2,\zb_2) 
     \prod_{j=3}^n\phi^{\epsilon_j}_{\Delta_j-1}(X|z_j,\zb_j).
\end{multline}
Here $\A$ is the term in the amplitude linear in the background; for three- and four-point scattering, this gives the leading contribution for generic arrangements on the celestial sphere. Letting $\tau = \sqrt{-X^2}$ such that $X = \tau\hat{x}$, setting $\beta = \sum_j(\Delta_j-1)$, and defining the phase factor $\mathcal{N}_n = \prod_{j=1}^n(-i\epsilon_j)^{\Delta_j-s_j}$, we can isolate the $\tau$ dependency into a single integral 
\begin{multline}
               \A(1^{\epsilon_1}_{\Delta_1,-}2^{\epsilon_2}_{\Delta_2,-}3^{\epsilon_3}_{\Delta_3,+}\cdots n^{\epsilon_n}_{\Delta_n,+}) = \frac{\mathcal{N}_n}{(2\pi)^42^{n-4}}\frac{z_{12}^3}{z_{23}\cdots z_{n1}}\int_{i\infty}^\infty d\tau\tau^{-1-\beta}\phi_B(\tau)\\
               \int \widetilde{d^3\hat{x}}\frac{\Gamma(\Delta_1+1)\Gamma(\Delta_2+1)}{(-q(z_1,\zb_1)\cdot\hat{x}-i\epsilon_1\epsilon/\tau)^{\Delta_1+1}(-q(z_2,\zb_2)\cdot\hat{x}-i\epsilon_2\epsilon/\tau)^{\Delta_2+1}} \prod_{j=3}^n\frac{\Gamma(\Delta_j-1)}{(-q(z_j,\zb_j)\cdot\hat{x}-i\epsilon_j\epsilon/\tau)^{\Delta_j-1}}
\end{multline}
Here $\widetilde{d^3\hat{x}}$ is the measure over the entire $\hat{x}^2 = -1$ slice. The contour in $\tau$ can be closed around the positive imaginary axis since $\phi_B(\tau)$ is exponentially damped for $\Im\tau > 0$, leaving a $\beta$ dependent prefactor and an integral over an $AdS_3/\mathbb{Z}$ slice of Klein space:
\begin{equation}
\label{eq:fptmhvint}
    \begin{split}
        \A &=  \frac{\mathcal{N}_n}{(2\pi)^42^{n-4}}\frac{z_{12}^3}{z_{23}\cdots z_{n1}}\Gamma(\Delta_1+1)\Gamma(\Delta_2+1)\prod_{j=3}^n\Gamma(\Delta_j-1) \\
        &\times \int \widetilde{d^3\hat{x}}(-q(z_1,\zb_1)\cdot \hat{x})^{-\Delta_1-1}(-q(z_2,\zb_2)\cdot \hat{x})^{-\Delta_2-1}\prod_{j=3}^n(-q(z_j,\zb_j)\cdot\hat{x})^{-\Delta_j+1} \\
        &\times \int_0^{i\infty}d\tau\tau^{-1-\beta}\phi_B(\tau)(e^{-2\pi i\beta}-1)
    \end{split}
\end{equation}
We identify $(-q(z,\zb) \cdot \hat{x})^{-\Delta}$ as the scalar bulk-to-boundary propagator of weight $\Delta$ on the $\tau^2 = 1$ slice of Klein space. The integral over $\widetilde{d^3\hat{x}}$ then gives an $n$-point scalar contact Witten diagram which have been studied in detail in many places including ~\cite{Chen:2017gwd, Hijano:2016DFunction}. The integral over $\tau$ can be evaluated analytically
\begin{equation}
   f(\beta) =  \int_0^{i\infty}d\tau\tau^{-1-\beta}\phi_B(\tau)(e^{2\pi i\beta}-1) = -\left(\frac{m}{2i}\right)^\beta \pi^2\Gamma\left(-1-\frac{\beta}{2}\right)\Gamma\left(-\frac{\beta}{2}\right)(e^{-2\pi i\beta}-1) 
\end{equation}
so that we can write the $3-$point function explicitly 
\begin{equation}
\label{eq:fptmhv}
\begin{split}
 \A(1^{\epsilon_1}_{\Delta_1,-}2^{\epsilon_2}_{\Delta_2,-}3^{\epsilon_3}_{\Delta_3,+}) &= -\frac{\mathcal{N}_3(1+(-1)^{-\beta})}{32\pi^3} \frac{z_{12}^3}{z_{23}z_{31}} f(\beta)\\
    &\times \frac{\Gamma\left(\frac{\Delta_1+\Delta_2-\Delta_3+3}{2}\right)\Gamma\left(\frac{\Delta_1-\Delta_2+\Delta_3-1}{2}\right)\Gamma\left(\frac{-\Delta_1+\Delta_2+\Delta_3-1}{2}\right)\Gamma\left(\frac{\Delta_1+\Delta_2+\Delta_3-1}{2}\right)}{|z_{12}|^{\Delta_1+\Delta_2-\Delta_3+3}|z_{23}|^{\Delta_2+\Delta_3-\Delta_1-1}|z_{13}|^{\Delta_1+\Delta_3-\Delta_2-1}}
\end{split}
\end{equation}
and the $4-$point function
\begin{equation}
\begin{split}    
    \A(1^{\epsilon_1}_{\Delta_1,-}2^{\epsilon_2}_{\Delta_2,-}3^{\epsilon_3}_{\Delta_3,+}4^{\epsilon_4}_{\Delta_4,+}) &= \frac{\mathcal{N}_4(1+(-1)^{-\beta})}{2(2\pi)^4}\frac{z_{12}^3}{z_{23}z_{34}z_{41}} \\
    &\times \Gamma(\Delta_1+1)\Gamma(\Delta_2+1)\Gamma(\Delta_3-1)\Gamma(\Delta_4-1)f(\beta) \\
    &\times D_{\Delta_1+1,\Delta_2+1,\Delta_3-1,\Delta_4-1}(z_1,z_2,z_3,z_4). 
    \end{split}
    \end{equation}
Here $D$ denotes the $D$-function, suitably continued to Lorentzian signature\footnote{A prefactor of $(1+(-1)^{-\beta})/2$ arises from our measure $\widetilde{d^3\hat{x}}$ and from integrating over the entire $\hat{x}^2 = -1$ slice of Klein space, rather than the region covered by coordinates analytically continued from the future-directed Euclidean AdS$_3$ slice of Minkowski space.} \cite{DHoker:1999kzh}
\begin{equation}
    \frac{1+(-1)^{-\sum_{j=1}^4\Delta_j}}{2}D_{\Delta_1,\Delta_2,\Delta_3,\Delta_4}(z_1,z_2,z_3,z_4) = \int \widetilde{d^3\hat{x}}\prod_{j=1}^4(-q(z_j,\zb_j)\cdot \hat{x})^{-\Delta_j},
\end{equation}
a widely-studied special function defined as the four-point contact scalar Witten diagram \cite{Hijano:2016DFunction}. It is exciting that Mellin amplitudes in the massive scalar background are proportional to standard boundary correlators enountered in AdS$_3$ holography, and brings together AdS$_3$ and flat holography.  

\section{Leading Soft Theorems and OPEs}

We now examine the soft structure of these celestial amplitudes at leading order in the perturbation.  Introducing the background scalar field breaks various symmetries  so it is unclear how much of the soft algebra will survive. In this section, we show that the leading conformally soft theorem for positive helicity gluons is undeformed, while the leading conformally soft theorem for negative helicity particles is no longer visible in the perturbed theory already in the four-point MHV amplitude.

Starting from the $4$-point MHV amplitude to linear order in the background with generic $z_i, \zb_i$,
\begin{equation}
\begin{split}
     \A(1^{\epsilon_1}_{\Delta_1,-}2^{\epsilon_2}_{\Delta_2,-}3^{\epsilon_3}_{\Delta_3,+}4^{\epsilon_4}_{\Delta_4,+}) &= \frac{\mathcal{N}_4}{(2\pi)^4}\frac{z_{12}^3}{z_{23}z_{34}z_{41}}\Gamma(\Delta_1+1)\Gamma(\Delta_2+1)\Gamma(\Delta_3-1)\Gamma(\Delta_4-1) f(\beta) \\
     &\times \int\widetilde{ d^3\hat{x}}\frac{1}{(-q_1\cdot \hat{x})^{\Delta_1+1}(-q_2\cdot\hat{x})^{\Delta_2+1}(-q_3\cdot\hat{x})^{\Delta_3-1}(-q_4\cdot\hat{x})^{\Delta_4-1}}
\end{split}
\end{equation}
we see that there is a simple pole where $\Delta_4 \to 1$. Cancelling the pole, we have that 
\begin{equation}
    \begin{split}
        \lim_{\Delta_4 \to 1}(\Delta_4-1) \A(1^{\epsilon_1}_{\Delta_1,-}2^{\epsilon_2}_{\Delta_2,-}3^{\epsilon_3}_{\Delta_3,+}4^{\epsilon_4}_{\Delta_4,+}) &= \frac{\mathcal{N}_3}{(2\pi)^4}\frac{z_{12}^3}{z_{23}z_{34}z_{41}}\Gamma(\Delta_1+1)\Gamma(\Delta_2+1)\Gamma(\Delta_3-1) f(\beta) \\
     &\times \int \widetilde{d^3\hat{x}}\frac{1}{(-q_1\cdot \hat{x})^{\Delta_1+1}(-q_2\cdot\hat{x})^{\Delta_2+1}(-q_3\cdot\hat{x})^{\Delta_3-1}} \\
     &= \frac{1}{2}\frac{z_{31}}{z_{34}z_{41}} \A(1^{\epsilon_1}_{\Delta_1,-}2^{\epsilon_2}_{\Delta_2,-}3^{\epsilon_3}_{\Delta_3,+})
    \end{split}
\end{equation}
matching the expected positive helicity soft theorem exactly. Conversely, in the case of a negative helicity soft mode the limit
\begin{equation}
    \lim_{\Delta_1\to 1}(\Delta_1-1)\A(1^{\epsilon_1}_{\Delta_1,-}2^{\epsilon_2}_{\Delta_2,-}3^{\epsilon_3}_{\Delta_3,+}4^{\epsilon_4}_{\Delta_4,+}) = 0
\end{equation}
is finite, in contrast with the celestial amplitude without background \cite{Pate:2019soft}. This is consistent with the vanishing of the amplitude $\A^\phi(1^-2^+3^+)$ at linear order. Since we are working with MHV amplitudes in a background where $\A^\phi(1^\pm2^+3^+)$ vanishes, it is necessary to work with NMHV and higher degree amplitudes to understand whether the negative-helicity conformally soft theorem is deformed.

%


Working in Klein space incoming and outgoing operators are related by a spacetime rotation \cite{Atanasov:2021} and their OPEs are closely related.\footnote{ Since  a point and its anitpode on the  celestial torus are on each other's light cones, OPE singulairities can occur between in and out modes.} Here, we compute only outgoing-outgoing OPEs. The background field couples chirally to the gauge field, that is, it couples only to the anti-self-dual field strength in \eqref{YM_dil_L}. Thus, we expect that the OPEs between positive helicity gluons are preserved, while OPEs involving negative helicity gluons might be modified. We examine the $z_{34} \to 0$, $z_{23} \to 0$, and $z_{12} \to 0$ limits of the celestial amplitude in Equation \ref{eq:fptmhv} to investigate the $++$, $+-$, and $--$ OPEs respectively.

Consider first the OPE between the two positive helicity particles given by the limit $z_{34}\rightarrow0$. Starting from the integral representation in Equation \ref{eq:fptmhvint} and working to leading order in $z_{34}$, we can rewrite the denominator as 
\begin{multline}
    \frac{1}{(-q(z_3,\zb_3)\cdot X)^{\Delta_3-1}(-q(z_4,\zb_4)\cdot \hat{x})^{\Delta_4-1}} = \sum_{m=0}^\infty \frac{\left(\Delta_3-1\right)_m}{\left(\Delta_3+\Delta_4-2\right)_m}\frac{\zb_{34}^m}{m!}\bpd_4^m\frac{1}{(-q(z_4,\zb_4)\cdot \hat{x})^{\Delta_3+\Delta_4-2}} + \O(z_{12})
\end{multline}
Plugging this expansion into Equation \ref{eq:fptmhvint} together with $\dfrac{z_{12}^3}{z_{23}z_{34}z_{41}} = \dfrac{1}{z_{34}}\dfrac{z_{12}^3}{z_{24}z_{41}} + O(z_{34}^0)$ gives the leading order collinear structure 
\begin{equation}
\begin{split}
    \A(1^-_{\Delta_1-}2^-_{\Delta_2-}3^+_{\Delta_3+}4^+_{\Delta_4+}) &=  \frac{1}{2z_{34}}\sum_{m=0}^\infty \frac{\Gamma(\Delta_3+m-1)\Gamma(\Delta_4-1)}{\Gamma(\Delta_3+\Delta_4+m-2)}\frac{\zb_{34}^m}{m!}\bpd_4^m \A(1^-_{\Delta_1-}2^-_{\Delta_2-}4^+_{\Delta_3+\Delta_4-1+}) + \O(z_{12}^0)
\end{split}    
\end{equation}
where we kept the tower of antiholomorphic descendants explicit. This leading OPE is the same as the leading OPE between positive helicity conformal primary gluons in color-ordered MHV amplitudes in a trivial background \cite{Pate:2019ope}. 

Following a similar computation, the leading term in the $z_3 \to z_2$ limit is
\begin{equation}
\A(1^-_{\Delta_1,-}2^+_{\Delta_2,-}3^+_{\Delta_3,+}4^-_{\Delta_4,+}) = \frac{1}{2z_{23}}\frac{\Gamma(\Delta_2+1)\Gamma(\Delta_3-1)}{\Gamma(\Delta_2+\Delta_3)} \mathcal{A}(1^-_{\Delta_1-}2^+_{\Delta_2+\Delta_3-1,-}4^-_{\Delta_4,+}) + \mathcal{O}(z_{23}^0,\zb_{23}).
\end{equation}
Which again agrees with the $+-$ collinear singularity in pure Yang-Mills MHV celestial amplitudes \cite{Pate:2019ope}.

By examining the leading $z_{12}$ and $\zb_{12}$ dependence of the four-point function, we see that the leading term describes a correlation function of two positive helicity gluons and an operator of weight $\Delta_1 + \Delta_2 + 3$ and spin $J = 1$. Due to the vanishing of the $\mathcal{A}^\phi(1^-2^+3^+)$ three-point function in this background, we see no contribution from a single gluon of negative helicity. Notably, an operator of the same weight gives the leading order contribution to the collinear limit between two negative helicity gluons in an $n$-point MHV amplitudes for $n > 4$.
%


To summarize, at leading order our MHV amplitude retain the same OPE limits and soft structure when gluons of positive helicity are concerned, but there is no negative helicity soft theorem and the OPE limit between gluons of negative helicity are deformed. It is interesting to note that the structure of the deformed OPEs we obtained is similar to the chiral algebra construction of \cite{Costello:2022wso} for the soft sector of self dual Yang-Mills and perturbative expansions around it. It would be interesting to see how our amplitude could be related to such expansions like the MHV formalism \cite{Cachazo:2004kj,Boels:2007qn}. We also note that for amplitudes with 5 or more particles the leading $--$ collinear limit of our celestial amplitude seems to match the flat space amplitude result. 
To investigate this further it would be very interesting to study the subleading terms in the OPEs and soft factors, and compare them to the terms obtained from the usual amplitudes \cite{Ebert:2020nqf}. Moving beyond the MHV section is another way to ascertain if the structure of the soft and collinear limits are deformed further.

\section*{Acknowledgements}

We are grateful to Kevin Costello, Indranil Halder, Natalie Paquette, Ana-Maria Raclariu, Atul Sharma, Tomasz Taylor, and Bin Zhu for useful conversations. This work was supported in part by DOE grant de-sc/0007870. WM gratefully acknowledges support from NSF GRFP grant DGE1745303. EC thanks the support of the Frankel-Goldfield-Valani Research Fund.

 \bibliographystyle{utphys}
 \bibliography{bib}

\end{document}